\pdfoutput=1

\documentclass[11pt]{article}
\usepackage{array}
\newcolumntype{M}[1]{>{\centering\arraybackslash}m{#1}}
\usepackage[final]{acl}

\usepackage{times}
\usepackage{latexsym}

\usepackage[T1]{fontenc}

\usepackage[utf8]{inputenc}

\usepackage{microtype}

\usepackage{inconsolata}

\usepackage{graphicx}

\usepackage{array}
\usepackage{float}
\usepackage{cite}
\usepackage{amsmath,amssymb,amsfonts}
\usepackage{algorithmic}
\usepackage{graphicx}
\usepackage{textcomp}
\usepackage{xcolor}
\usepackage{array}
\usepackage{graphicx}
\usepackage{caption}
\usepackage{subcaption}

\title{Exploring Emotion-Sensitive LLM-Based Conversational AI}

\author{Antonin Brun, {\bf Ruying Liu},  \\
        {\bf Aryan Shukla,} {\bf Frances Watson} \\
        \textit{University of Southern California} \\
        Los Angeles, CA, USA\\
        \texttt{\{abrun,ruyingli,aryanshu,fwatson\}@usc.edu} \\ 
        \And
        Jonathan Gratch\\
        \textit{Institute for Creative Technologies} \\
        \textit{University of Southern California}\\
        Los Angeles, CA, USA\\
        \texttt{gratch@ict.usc.edu}}

\begin{document}
\maketitle
\begin{abstract}
Conversational AI chatbots have become increasingly common within the customer service industry. Despite improvements in their emotional development, they often lack the authenticity of real customer service interactions or the competence of service providers. By comparing emotion-sensitive and emotion-insensitive LLM-based chatbots across 30 participants, we aim to explore how emotional sensitivity in chatbots influences perceived competence and overall customer satisfaction in service interactions. Additionally, we employ sentiment analysis techniques to analyze and interpret the emotional content of user inputs. We highlight that perceptions of chatbot trustworthiness and competence were higher in the case of the emotion-sensitive chatbot, even if issue resolution rates were not affected. We discuss implications of improved user satisfaction from emotion-sensitive chatbots and potential applications in support services.
\end{abstract}

\section{Introduction}
Within the context of customer service, emotional labor involves employees managing their feelings and expressions to promote customer satisfaction \citep{hochschild_emotion_1979}. Beyond the actual service provided, customers are more likely to trust an employee or organization when their emotional needs are met, particularly if these emotions are perceived as genuine. This understanding has motivated increased interest in embedding emotional-sensitivity into customer service chatbots \citep{bilquise_emotionally_2022, li_towards_2021, pamungkas_emotionally-aware_2019}. However, it is unclear if emotions from an AI would have the same effect. For example, a recent study found that emotion-sensitivity backfired when customers were aware that it was generated by AI, leading them to assume it was inauthentic \citep{han_bots_2023}. Yet the growing sophistication of large language models (LLMs) suggests that these systems can convey emotionally nuanced language that might be perceived as authentic. 

To examine this, we performed a study using LLMs to simulate an IT customer service interaction, manipulating the emotional sensitivity of the model. Our preliminary findings indicate that while the addition of emotional sensitivity did not alter the problem-solving abilities of the system (two-thirds of participants reported having their issue resolved and this was the same across both conditions), it did significantly increase their trust and belief in the competence of the emotionally-sensitive system. These results align with findings from emotional labor and reinforce attempts to integrate emotion into such systems. However, they also raise concerns about users potentially over-trusting systems that exhibit human-like emotional expressions \citep{hancock_ai-mediated_2020}. Future work will need to investigate if user trust is appropriately calibrated when interacting with emotionally-sensitive customer service agents (e.g., do people under-trust non-emotional agents or over-trust emotional ones?) and examine the longer-term consequences of such technology for both customers and employees.

\section{Background}
Empathy has long been recognized as a key factor in enhancing customer satisfaction within the service industry \citep{kernbach_impact_2005}. Well-tailored emotional responses from a customer service agent may help regulate the emotion of an impassioned customer \citep{delcourt_effects_2013}. Early evidence suggests that chatbots may serve to benefit from increased emotional sensitivity in the same way that humans do, although this relationship has not yet been studied comprehensively \citep{gelbrich_emotional_2021}. As a novel technology, AI chatbot systems must also build trustworthiness to increase user satisfaction and perceived competence  \citep{huang_can_2024}. 

Given this context, it is important to understand whether emotional sensitivity in chatbots enhances users' perceptions of competence, supportiveness, and overall pleasantness in customer service interactions. Our study aims to explore the relationship between emotional sensitivity and quality of customer service. To access emotional sensitivity, we employ sentiment analysis techniques such as VADER and use the current state-of-the-art LLM chatbots to simulate emotionally nuanced interactions \citep{hutto_vader_2014, mehmood_enhanced_2020}.

\section{Experiment}
\subsection{Chatbot Design}
\begin{figure}[htbp]
    \centerline{\includegraphics[width=\columnwidth]{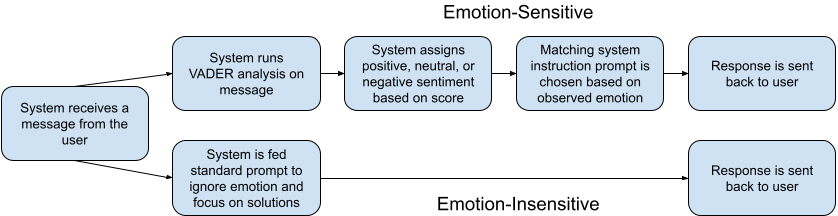}}
    \caption{Emotion-sensitive and -insensitive prompt engineering logic.}
    \label{fig:1}
\end{figure}

For our study, we decided to stray from the traditional, rule-based IT chatbot systems in favor of an LLM base to allow for improved implementation of emotional communication, using ChatGPT-3.5 (GPT). While one way to have an emotion-sensitive chatbot subject would be to simply allow GPT to detect emotional language in the user’s prompt and respond accordingly based on its training, we opted for a more controlled experimental design where the model was instructed to follow standard tactics used in customer service to address the customer’s emotion \citep{magids_new_2015}. Our design recognized the customer’s emotions using a previously-validated model \citep{hutto_vader_2014}, and then gave GPT specific prompts on how to respond based what was recognized. We further use VADER to assess GPT’s output to see if it conforms with these prompts for the sake of objectivity and consistency. The following diagram in figure \ref{fig:1} below describes the operation of the emotion-sensitive chatbot as well as the emotionally insensitive control.

In essence, in the case of the emotion-sensitive system, we ran an analysis on each user input to select from three different system prompts that would specify a emotion and an appropriate response tone to GPT based on the VADER score. In the case of the emotion-insensitive system, the system was always instructed to remain stoic and problem-focused. With this experimental structure, we believe that the user’s emotion could be experimentally measured, and the emotionality of the bot’s responses could be controlled within the functional bounds of GPT. 

\subsection{Study Design, Participants and Measures}
This study was conducted with 30 participants recruited over personal communication channels, principally consisting of friends and family.
To evaluate our experiment, we used a survey composed of four different sections. First, we provided participants with one example of an IT issue (out of 7 scenarios) and a current emotional state (out of 6 states). For example, the users could be prompted with the following IT issue: “Every time you try to open Word, the app crashes immediately without any error”, and emotional state: “You have had a rough day at work, and you cannot wait to be back home”. These scenarios and emotions were meant to contextualize the user’s interaction with the bot, foster genuine interactions with the chatbot by establishing a hypothetical situation, and encourage longer interactions. Second, we assessed the participants’ emotions in the fictitious situation. We used 20 questions following the PANAS scale, rated on a 5-point likert scale \citep{watson_development_1988}. The PANAS scale consists of two mood scales: positive affect and negative affect, each of which are further divided into ten addiitonal affective items. Positive and negative emotions are calculated by averaging the ratings of the 10 positive and 10 negative items, respectively. 
Third, we directed the participants to interact with one of the two chatbots selected at random. Finally, we aimed to assess the participants' emotions after completing the experiment (i.e., interaction with a chatbot) using the same questions described in the second part of the survey. We also evaluated the quality of their interaction using the following questions: “Was your concern or problem resolved after using the chatbot? [Yes/No]”, “What did you like most about your interaction with the chatbot?”, “What did you like least about your interaction with the chatbot?”, and “Do you have any other comments/concerns/questions you would like to discuss?”. 

\section{Results}

\subsection{Manipulation Check}
To validate our experiment, it was important to consider the potential for GPT to exhibit an unintended emotional response beyond the bounds of our experiment. As a manipulation check, we ran a supplemental analysis on GPT’s responses to user queries to ensure that the chatbot was behaving as expected by comparing the VADER scores of incoming user messages to the VADER score of both chatbot’s output. We considered a margin of \textpm 0.1 from zero points on the VADER score to be neutral, with scores further in the positive direction being considered to have a “Positive” sentiment and scores further in the negative direction being considered to have a "Negative” sentiment.

\begin{figure*}[ht]
    \centerline{\includegraphics[width=\linewidth]{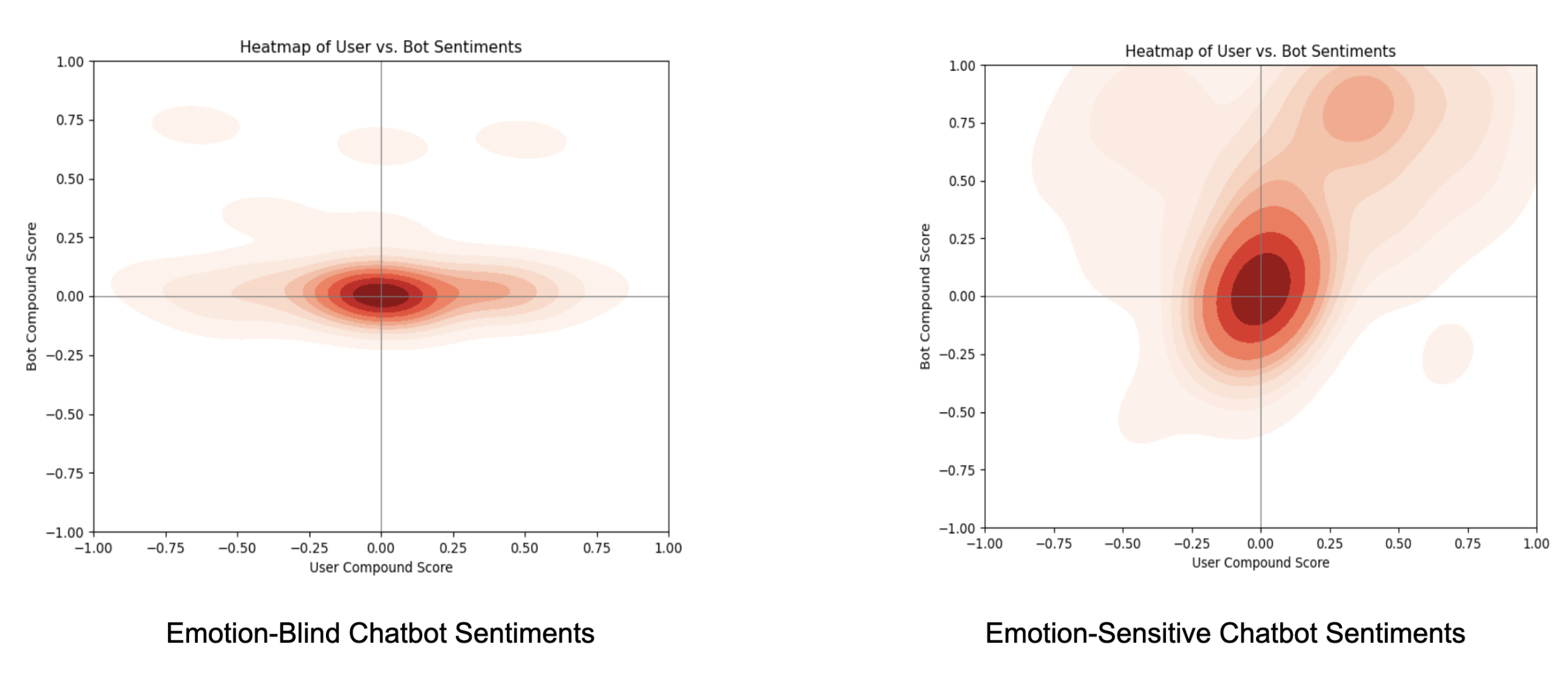}}
    \caption{Heatmaps of VADER-measured sentiments of user message-bot response pairs for both emotion-insensitive (left) and emotion-sensitive (right) chatbots.}
    \label{fig:2}
\end{figure*}


As shown in figure \ref{fig:2}, the emotion-insensitive bot’s sentiments stayed close to zero the vast majority of the time, whereas the emotion-sensitive bot’s sentiments were far more variable. In the case of the emotionless bot, 63 out of 68 (92.6\%) input prompts collected throughout the experiment resulted in a bot response that had a near-zero sentiment score. For the emotional bot, 82 out of 91 (90.1\%) responses accurately mirrored the non-neutrality of the relevant user input. In both cases, we were able to confirm that the model was behaving as expected, and we felt comfortable moving forward. 

\subsection{Problem-Solving Performance}
Amongst the 30 participants recruited, 14 were randomly assigned to interact with the emotionally sensitive chatbot, and 16 interacted with the emotionally insensitive chatbot. Regarding the question "Was your concern or problem resolved after using the chatbot?", our Chi-square analysis did not reveal any significant differences between groups ($\chi^2$ (1, N = 30) = 0.268, p = 0.605), indicating that adding emotion did not alter the problem-solving abilities of the system. Overall 67\% participants were satisfied with chatbots' performance in resolving users’ concerns.

The equivalence of problem-solving performance between the two groups is also supported by users' change in negative emotions. We used mixed ANOVAs to examine the overall effects across time relative to the negative emotion measured using the PANAS scale. We found a significant effect of time (i.e., pre-interaction and post-interaction) on negative emotion (F (1, 28) = 25.086, p $<$ 0.001, partial $\eta^2$ = 0.473) where negative emotion significantly decreased after interacting with either chatbot (M$_{pre}$= 2.486, M$_{post}$ = 1.898). However, we did not find any significant effect or interaction of group (i.e., emotional and non-emotional) (p $>$ 0.05). These results indicate that interacting with either chatbot may help reduce negative emotions.

\subsection{User Impressions}
We conducted one-way ANOVAs with two groups (i.e., emotional and unemotional) as the between-subjects variable to study the effect of emotion-awareness of a chatbot on user feedback. We found significant differences, as shown in Appendix \ref{tab1:appendix}, where participants who interacted with the emotionally sensitive chatbot reported higher agreement on the chatbot’s capability, knowledge, trustworthiness, being supportive and understanding, and willingness to use it in the future. Overall, the results show that an emotionally sensitive chatbot provides a better user experience than an emotionally insensitive chatbot.

In addition to the quantitative results, we also looked into participants' responses to the open-ended question "What did you like most about your interaction with the chatbot?". The emotional group seems to have perceived more on the social and emotional aspects of their interactions with the chatbot by mentioning phrases such as  "felt like conversation", "friendly", "very kind", and "had a great personality". This suggests that users perceived the emotionally sensitive chatbot to be personable and engaging. 

\section{Discussions}


Our analysis indicated that both chatbots have equivalent problem-solving performance, with two-thirds of participants reporting their problem solved in both cases. The analysis of emotional states using the PANAS scale revealed a significant decrease in negative emotions following interactions with either chatbot. However, the two chatbots performance began to deviate in terms of user perceptions and experiences. Participants who interacted with the emotionally sensitive chatbot scored the bot higher in terms of capability, knowledge, trustworthiness, supportiveness, understanding, and willingness to use it in the future when compared to participants who interacted with the emotionally insensitive bot. While both chatbots were equally effective in resolving users' concerns, participants in the emotional group tended to perceive their interactions in more socially and emotionally engaging terms, which aligns with theories that imbuing technology with emotional sensitivity can enhance user acceptance and satisfaction by creating more human-like interactions \citep{rapp_human_2021}.

Our results demonstrated statistically significant findings regarding our second hypothesis, that users would rate the emotion-sensitive chatbot at a higher competence level than the emotion insensitive chatbot. However, our experiment revealed no significant differences in regards to our first hypothesis, suggesting that the two chatbots did not impact the positivity of the user's emotions differently. Thus, while users did not mirror the bot's emotional state, they did tend to associate emotional sensitivity with competence. In the customer service field, these findings demonstrate that emotional receptiveness of a customer service agent is an important component of increasing satisfaction, especially in the often high-pressure setting of customer service support.

\section{Limitations and Future Work}

There are several points to note regarding the validity and future directions of this study. Firstly, the sample size of the study was extraordinarily small; larger studies would be encouraged to improve the validity of results both in terms of perceived chatbot competence and emotional impact. The participant pool was also limited to relatives and close friends, which all might have similar interactions with customer support services and AI chatbots in general. We hypothesize that since we prompted participants with fictitious scenarios (that our team generated), this could have had an effect on the participants' interaction with the chatbot and their perceived emotions.

Additionally, our experimental design, which introduced fictitious scenarios, may have influenced participants' interactions and emotional responses towards the chatbots. Future studies could explore alternative scenario designs to minimize potential biases and improve consistency. Experimenting with different prompts could further elucidate their impact on user receptiveness.

Methodologically, our emotional sentiment analysis was based on single user messages, limiting the contextual understanding of emotional dynamics. In future experiments, sentiment analysis can be run using the context of entire conversations to better understand the impact that emotional context has on user emotion. 
We also acknowledge the limitations associated with our use of VADER for sentiment analysis. While VADER was efficient for general sentiment analysis, its rule-based nature may have overlooked nuanced emotional expressions crucial to our study. Given the topic's alignment with affective computing, the reliance on VADER could be perceived as a straightforward approach that ensures objectivity, but might not delve deeply into the subtleties of affective states. Future research could benefit from more sophisticated models like BERT-based approaches \citep{devlin2018bert} to account for the contextual and semantic intricacies of the data, and therefore capture subtler emotional nuances effectively.


\section{Conclusion}
By conducting a between-subjects study to compare user sentiments throughout interactions with an emotion-sensitive chatbot and an emotion-insensitive chatbot, we were able to find that perceptions of chatbot trustworthiness and competence were higher in the case of the emotion-sensitive chatbot. Still, we discovered that there was no significant difference between emotional states following interactions with either chatbot, suggesting that emotional sensitivity may prove to be more useful for competence ratings than emotional management in an IT context. In future studies, the level of this impact may be examined by controlling for variables such as prompts, IT scenarios, conversation context, and more. Future studies may also be able to better contextualize these results with experiments that integrate real-life IT scenarios rather than simulated ones. With the discovery that emotional sensitivity may be perceived to make a digital chatbot system more competent, customer service support systems may be able to improve user satisfaction. 

\section*{Ethics Statement}

This pilot study was performed as part of a class project in an Affective Computing course and is intended to lay the groundwork for more rigorous future evaluations. The experimental design conforms to a protocol that was classified as exempt by our university’s ethical board (i.e., involves minimal risks). Participants were a convenience sample of friends and family members who were well-acquainted with the research but blind in advance to the nature of the study. As such, the diversity of the sample was limited, and our sample may not be representative of typical users of customer support technology due to biases. Thus, readers should take caution when generalizing the implications of these findings to real-world contexts. 
 
Finally, we acknowledge that the use of emotional AI in customer service applications could be viewed as manipulative. Real-world applications of this technology should consult ACM’s Code of Ethics and Professional Conduct and be sensitive to user autonomy. Users should be allowed to opt into the use of emotional behaviors. Future research should carefully consider the ethical implications of this technology in for-profit applications.

\nocite{*}
\bibliography{References}

\appendix

\section{User interaction ratings.}
\label{tab1:appendix}

\begin{table*}[htbp]
    \begin{tabular}{>{\centering\arraybackslash}m{0.25\textwidth} >{\centering\arraybackslash}m{0.25\textwidth} >{\centering\arraybackslash}m{0.25\textwidth} >{\centering\arraybackslash}m{0.25\textwidth}} 
        \hline 
        \textbf{Statement (5-point Likert-scale)}   & \textbf{Statistical results}  & \textbf{Emotion-sensitive chatbot}    & \textbf{Emotion-insensitive chatbot} \\
        \hline
        The chatbot is capable of handling complex queries. & F(1, 28) = 4.473 \newline \textbf{p = 0.043} \newline Partial $\eta^2$ = 0.138 & M = 4.286 \newline Std. = 0.726 & M = 3.500 \newline Std. = 1.211 \\
        \hline
        I believe the chatbot has the necessary knowledge to answer my questions. & F(1, 28) = 7.637 \newline \textbf{p = 0.010} \newline Partial $\eta^2$ = 0.214 & M = 4.357 \newline Std. = 0.745 & M = 3.438 \newline Std. = 1.031 \\
        \hline
        Overall I trust this chatbot to assist me with my needs. & F(1, 28) = 8.506 \newline \textbf{p = 0.007} \newline Partial $\eta^2$ = 0.233 & M = 4.143 \newline Std. = 0.770 & M = 3.125 \newline Std. = 1.088 \\
        \hline
        I would use this chatbot again in the future for similar inquiries. & F(1, 28) = 7.433 \newline \textbf{p = 0.011} \newline Partial $\eta^2$ = 0.210 & M = 4.357 \newline Std. = 0.280 & M = 3.313 \newline Std. = 0.262 \\
        \hline
        I feel that the chatbot was supportive and understanding. & F(1, 28) = 4.068 \newline \textbf{p = 0.053} \newline Partial $\eta^2$ = 0.127 & M = 4.214 \newline Std. = 0.975 & M = 3.313 \newline Std. = 1.401 \\
        \hline
    \end{tabular}
\end{table*}

\end{document}